\documentclass[11pt]{cernrep}
\usepackage{graphicx}
\usepackage{cite}
\long\def\symbolfootnote[#1]#2{\begingroup%
     \def\thefootnote{\fnsymbol{footnote}}\footnote[#1]{#2}\endgroup} 

\begin{document}
 \title{\vspace*{-.3in}
\flushright{\rm \small UCD-2004-18 \\[-2pt]LBNL-55457\\[-2pt]CERN-PH-TH/2004-018\\[-2pt]  MCTP-04-05 \\[-8pt] hep-ph/0402062} \\\begin{center}{THE INVISIBLE HIGGS DECAY WIDTH \\[1pt] 
IN THE  ADD MODEL AT THE LHC~\symbolfootnote[2]{~~~~To appear in the Proceedings of the Les Houches Workshop 2003: ``Physics at TeV Colliders'', ed. F. Boudjema}}\end{center}}
\author{M. Battaglia$^1$, D. Dominici$^2$\symbolfootnote[3]{~~~~On leave from
Dipartimento di Fisica, Univ. Firenze, Sesto F. (FI) 50019, Italy}~, J.F. Gunion$^3$ and J.D. Wells$^4$}
\institute{$^1$Department of Physics, University of California
at Berkeley, and Lawrence Berkeley National Laboratory,
Berkeley, CA 94720; 
$^2$
 Department of Physics, CERN, Theory Division, CH-1211 Geneva 23, Switzerland; $^3$Department of Physics,
University of California at Davis, Davis, CA 95616; 
$^4$Department of Physics, University of Michigan, Ann Arbor, MI 48109-1120}
\maketitle
%
\def\to{\rightarrow}
\def\ptl{\partial}
\def\beq{\begin{equation}}
\def\eeq{\end{equation}}
\def\bea{\begin{eqnarray}}
\def\eea{\end{eqnarray}}
\def\nn{\nonumber}
\def\half{{1\over 2}}
\def\rhalf{{1\over \sqrt 2}}
\def\calo{{\cal O}}
\def\cala{{\cal A}}
\def\call{{\cal L}}
\def\calm{{\cal M}}
\def\del{\delta}
\def\eps{\epsilon}
\def\lam{\lambda}
\def\anti{\overline}
\def\delfac{\sqrt{{2(\del-1)\over 3(\del+2)}}}
\def\heff{h'}
\def\square{\boxxit{0.4pt}{\fillboxx{7pt}{7pt}}\hspace*{1pt}}
    \def\boxxit#1#2{\vbox{\hrule height #1 \hbox {\vrule width #1
             \vbox{#2}\vrule width #1 }\hrule height #1 } }
    \def\fillboxx#1#2{\hbox to #1{\vbox to #2{\vfil}\hfil}   }

\def\gev{~{\rm GeV}}
\def\mev{~{\rm MeV}}
\def\gam{\gamma}
\def\sn{s_{\vec n}}
\def\sm{s_{\vec m}}
\def\mm{m_{\vec m}}
\def\mn{m_{\vec n}}
\def\mh{m_h}
\def\sumn{\sum_{\vec n>0}}
\def\summ{\sum_{\vec m>0}}
\def\vl{\vec l}
\def\vk{\vec k}
\def\ml{m_{\vl}}
\def\mk{m_{\vk}}

\centerline{\bf Abstract}

\noindent 
Assuming flat universal extra dimensions,
we demonstrate that for a light Higgs boson 
the process $pp\to W^*W^* +X \to Higgs,graviscalars +X \to invisible+X$
will be observable at the $5~\sigma$ level at the LHC 
for the 
portion of the Higgs-graviscalar mixing ($\xi$) and effective
Planck mass ($M_D$) parameter space where channels
relying on visible Higgs decays fail to achieve
a $5~\sigma$ signal.  Further, we show that 
even for very modest values of $\xi$ 
the invisible decay signal 
probes to higher $M_D$ than does the ($\xi$-independent) 
jets/$\gam$ + missing energy signal
from graviton radiation.  We also discuss various
effects, such as Higgs decay to two graviscalars, that
could become important when $\mh/M_D$ is of order 1.

\vspace*{-.15in}
\section{INTRODUCTION}
In several  extensions of the Standard Model (SM) there
exist mechanisms which modify the Higgs production/decay 
rates in channels that are observable at the LHC.
One example is the Randall Sundrum model
where the   Higgs-radion mixing not only gives detectable 
reductions (or enhancements) in Higgs 
yields, but also allows the possibility of direct 
observation of radion production and 
decay~\cite{Dominici:2002jv,Battaglia:2003gb}. 
It is also possible for the Higgs rate 
in visible channels to be reduced
as a result of a substantial invisible width.
For example, this occurs in supersymmetric
models when the Higgs has a large 
branching ratio into the lightest gravitinos or neutralinos.
Invisible decay of the Higgs is also predicted 
in models with large extra dimensions felt by gravity (ADD)
\cite{Arkani-Hamed:1998rs,Antoniadis:1998ig}. 
In ADD models the presence of an
interaction between the Higgs $H$ and the 
Ricci scalar curvature of the induced 4-dimensional metric $g_{ind}$, 
generates, after the usual shift $H=({v+ h\over \sqrt{2}},0)$,
the following mixing term \cite{Giudice:2000av}
\begin{equation}
{\cal L}_{\rm mix}=\epsilon  h \sum_{\vec n >0}s_{\vec n}
\end{equation}
with
\beq
\eps=-{2\sqrt 2\over M_P}\xi v \mh^2\sqrt{{3(\del-1)\over \del+2}}\,.
\eeq
Above, $M_P=(8\pi G_N)^{-1/2}$ is the Planck mass, $\delta$ is the number of extra
dimensions, $\xi$ is a dimensionless parameter and
$s_{\vec n}$ is a graviscalar KK excitation with
mass $m_{\vec n}^2=4\pi^2 \vec n^2/L^2$, $L$ being the
size of each of the extra dimensions.
(Note that with respect to \cite{Giudice:2000av}
our normalization is such that we have taken
only the real part of the $\phi_G^{\vec n}$ fields,
writing $\phi_G^{\vec n}={1\over\sqrt 2}(s_{\vec n}+i a_{\vec n})$ 
and using $\phi_G^{\vec n}=[\phi_G^{-\vec n}]^*$
to restrict  sums to $\vec n>0$, by which we mean the first non-zero entry
of $\vec n$ is positive.) After diagonalization of the full mass-squared matrix
one finds that the physical eigenstate, $h'$,  acquires
admixtures of the graviscalar states and vice versa.
Dropping $\calo(\eps^2)$ terms and higher,
\beq
h'\sim \left[h-\sum_{\vec m>0}{\eps\over \mh^2-i \mh
\Gamma_{h}-m_{\vec m}^2}s_{\vec m}\right]\,,\quad
s'_{\vec m}\sim \left[ s_{\vec m}+{\eps\over \mh^2-i\mh\Gamma_{h} 
-m_{\vec m}^2} h\right]\,.
\label{eigenstate}
\eeq
In computing a process such as $WW\to h'+\sum_{\vec m>0}s_{\vec m}' \to F$,
normalization and admixture corrections of order $\eps^2$ that 
are present must be taken into account and the full coherent sum
over physical states must be performed. The result at the
amplitude level is 
\beq
\cala(WW\to F)(p^2)\sim {g_{WWh}g_{h F}\over  
p^2-\mh^2+i\mh\Gamma_h+iG(p^2)+F(p^2)}
\label{amplitude}
\eeq
where 
$F(p^2)\equiv -\eps^2 {\rm Re} \left[\sum_{\vec m>0}{1\over
    p^2-m_{\vec m}^2}\right]$ and $G(p^2)\equiv
-\eps^2{\rm Im}\left[\sum_{\vec m>0}{1\over p^2-m_{\vec m}^2}\right]$.
Taking the amplitude squared and
integrating over $dp^2$ in the narrow width approximation
gives the result
\beq
\sigma(WW\to h'+\sum_{\vec m>0}s'_{\vec m}\to F)=\sigma_{SM}(WW\to h \to F)
\left[{1\over 1+F'(m_{h\,ren}^2)}\right]\left[{\Gamma_h\over \Gamma_h+\Gamma_{h\to graviscalar}}\right]
\label{xsec}
\eeq
where $m_{h\,ren}^2-\mh^2+F(m_{h\,ren}^2)=0$ and we have defined
$\mh\Gamma_{h\to graviscalar}\equiv G(m_{h\,ren}^2)$. We will argue
that for a light Higgs boson both the wave function renormalization
and the mass renormalization effects will be small. In this case, 
the coherently summed
amplitude gives the Standard Model cross section suppressed by the
ratio of the SM Higgs width to the sum of the SM Higgs width and the
Higgs width arising from mixing with the graviscalars.

\section{INVISIBLE WIDTH}
As described, there is a decay of the Higgs arising from
the mixing (or oscillation) 
of the Higgs itself into the closest KK graviscalar
levels. These graviscalars are invisible since they
are weakly interacting and mainly reside in the 
extra dimensions whereas the Higgs resides on the brane.
The mixing width 
$\Gamma_{h\to graviscalar}\sim G(\mh^2)/\mh$ thus corresponds to an invisible decay 
width. The equation for $G(\mh^2)$ 
below eq.~(\ref{amplitude}) 
shows that it is calculated by extracting the imaginary
part of the mixing contribution to the Higgs self energy.
The result is \cite{Giudice:2000av,Wells:2002gq}
\begin{eqnarray}
\Gamma(h\to graviscalar)&\equiv& \Gamma(h\to 
\sum_{\vec n>0}s_{\vec n})=
2\pi\xi^2 v^2 \frac {3(\delta -1)}
{\delta +2}
\frac {m_h^{1+\delta}}{M_D^{2+\delta}}{S_{\delta -1}}\nn\\
&\sim& (16\,MeV) 20^{\delta -2} \xi^2
S_{\delta-1}\frac {3(\delta -1)}
{\delta +2} \left ( \frac {m_h}{150\, GeV} \right )^{1+\delta}
\left ( \frac 
{3\, TeV} {M_D}\right )^{2+\delta}
\label{invwidth}
\end{eqnarray}
where $S_{\delta-1}=2\pi^{\delta/2}/\Gamma(\delta/2)$ denotes the
surface of a unit radius sphere in $\delta$ dimensions while $M_D$ is
related to the $D$ dimensional reduced Planck constant ${\overline
  M}_D$ by $M_D= (2\pi)^{\delta/(2+\delta)}{\overline M}_D$.  Our eqs.
(\ref{invwidth}) are a factor of 2 larger than those presented in
refs.~\cite{Giudice:2000av,Wells:2002gq}.

\subsection{The wave function renormalization factor and mass renormalization}

A simple estimate of the quantity $F'(m_{h\,ren}^2)$, appearing in the
wave function renormalization factor found in eq.~(\ref{xsec}),
suggests that it is of order $\xi^2{\mh^4\over \Lambda^4}$, where
$\Lambda$ is an unknown ultraviolet cutoff energy presumably of order
$\Lambda\sim M_D$ \cite{Giudice:1998ck}. Assuming this to be the case, $F'$ will
provide a correction to coherently computed LHC production cross
sections that is  very probably quite small for the
$\mh\ll M_D$ cases that we are about to explore.  However, one must
keep in mind that a precise calculation of $F'$ is not possible.
Similarly, the mass renormalization from $F(m_{h\,ren}^2)$ should be
of order $\xi^2\mh^6/M_D^4$ and, therefore, small for $\mh\ll M_D$.
There are other incomputable sources of $v^4/M_D^4$ corrections
lurking in the theory beyond these sources, and the results presented
here are computed using the first, and perhaps only, calculable terms
in the perturbation series.

\subsection{Contribution to the invisible width from direct
two graviscalar decay}  

In addition to decay by mixing, one expects also a contribution to the
invisible width of the Higgs from its decays into two graviscalars.
This can be evaluated by using the transformation of
eq.~(\ref{eigenstate}) between the physical eigenstate $h'$ and the
unmixed $h$ to derive the relevant trilinear $h' s_k s_l$
vertices. These are used to compute the corresponding matrix element.
The final expression for $\Gamma(h'\to graviscalar~pairs)$ can be
written as
\begin{eqnarray}
\Gamma(h'\to graviscalar~pairs)
&=& {18\over \pi} {m_h^{3+2\del}v^2\over M_D^{4+2\del}}\xi^4
\left({\del-1\over\del+2}\right)^2 \left[{\pi^{\del/2}\over \Gamma(\del/2)}\right]^2 I\,,
\end{eqnarray}
where $I$ is an integral coming from the sum over all the possible
kinematically allowed $h'\to s_k s_l$ decays.  The integral $I$
decreases rapidly as $\del$ increases.  As a result, $\Gamma(h'\to
graviscalar~pairs)$ is only significant compared to $\Gamma(h\to
graviscalar)$ if $\del\leq 4$.  The ratio of the two widths is given
by:
\begin{equation}
{\Gamma(h'\to graviscalar~pairs)\over \Gamma(h\to graviscalar)}=
 {3(\del -1)\over 2\pi^2(\del+2)}\xi^2 \left({m_h\over M_D}\right)^{2+\del}
{\pi^{\del/2}\over \Gamma(\del/2)}I\,.
\end{equation}
From this result, we immediately see that even for small $\del$ the
pair invisible width will be smaller than the mixing invisible width
unless $m_h$ is comparable to $M_D$. 

To lowest order in $\xi^2(\mh/M_D)^{2+\del}$, 
decays of other states nearly degenerate with the $h'$
can be neglected in the computation of a 
cross section obtained by coherently summing over the 
$h'$ and the nearly degenerate $s'_{\vec m}$ states.
Thus, to this same order of approximation,
$\Gamma(h'\to graviscalar~pairs)$ should simply be
added to $\Gamma(h\to graviscalar)$ in the expression for
the narrow-width cross section of eq.~(\ref{xsec}). 

\begin{figure}[htbp]
\begin{center}
\includegraphics[width=7.0cm,height=5.0cm]{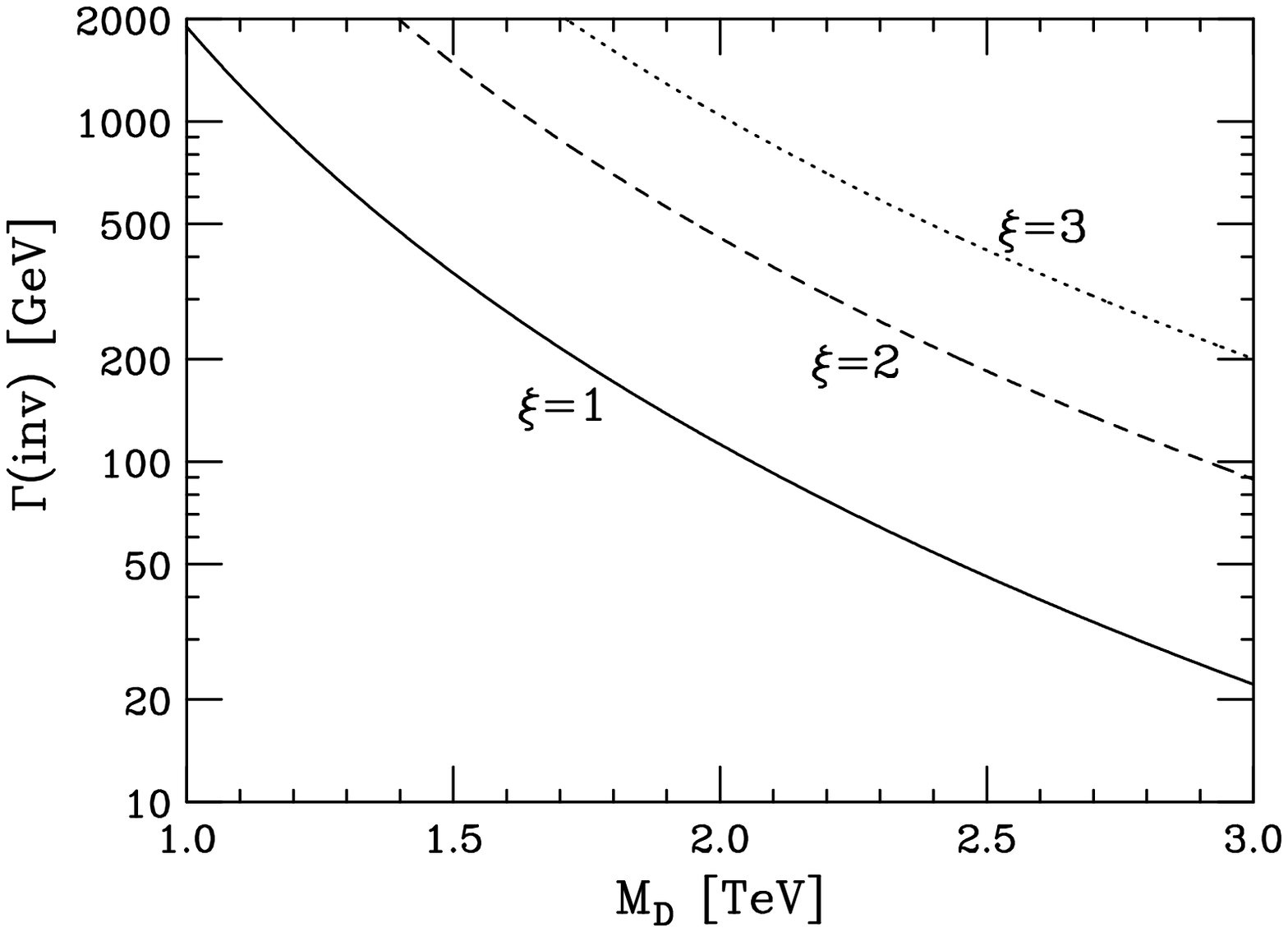} 
\includegraphics[width=7.0cm,height=5.0cm]{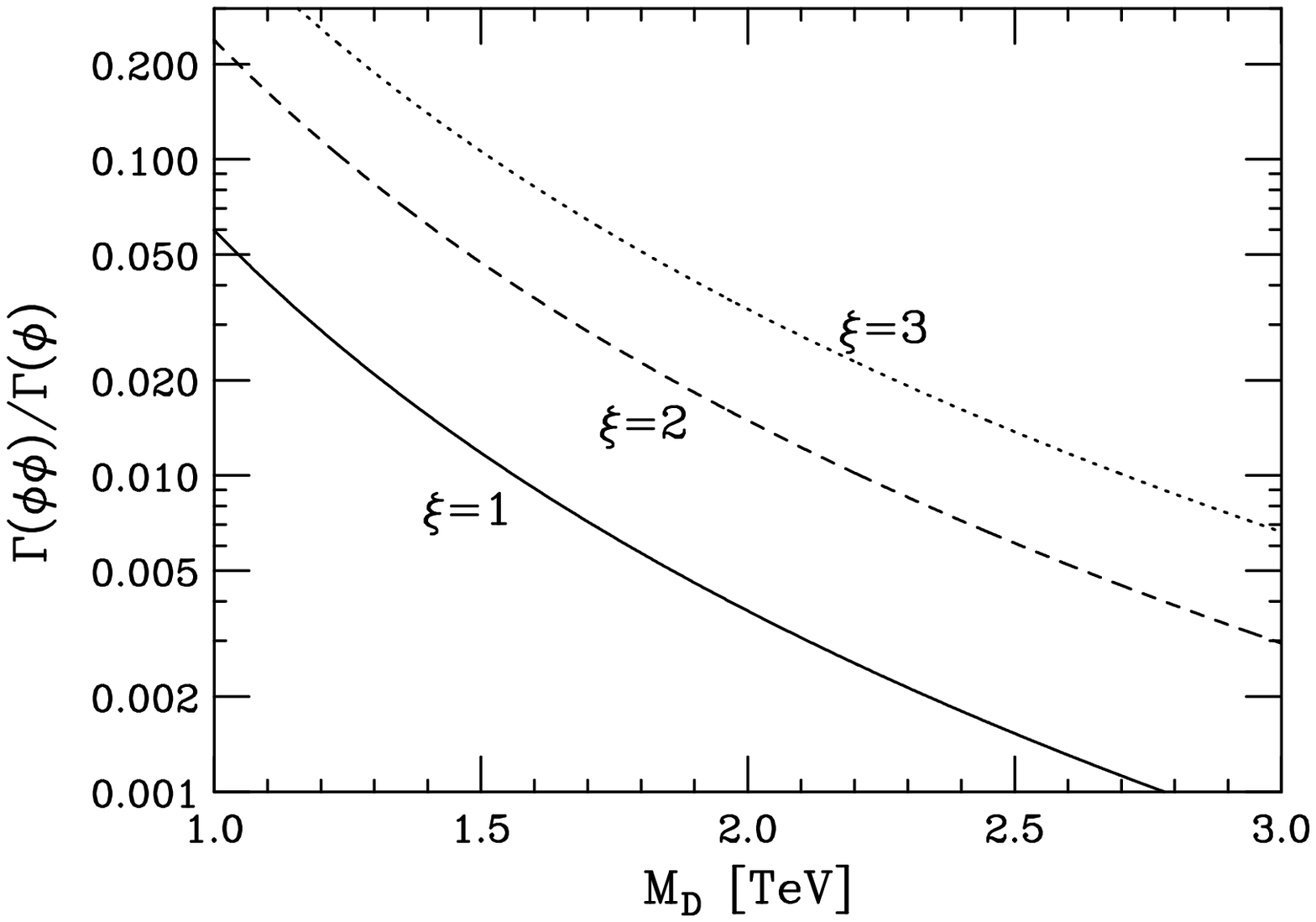}
\caption{In the left-hand plot, we display the total invisible width of a
  1~TeV Higgs boson into one and two graviscalars as a function of
  $M_D$ for various values of $\xi$ ($\xi=1$ solid, $\xi=2$ dashed,
  $\xi=3$ dotted).  For this plot we have fixed $\del=2$. The plot on
  the right shows the ratio of the two-graviscalars decay width 
  to the one-graviscalar decay width for the same choices of
  parameters.}
\label{paircomp}
\end{center}
\end{figure}
In Figure~\ref{paircomp}, we show an extreme case corresponding to
$\del=2$ and $m_h=1000\gev$. Depending on the values of the parameters
 $\xi$ and $M_D$, the pair
invisible width can be a significant correction to the invisible width
from direct mixing.  More generally, for $m_h>M_D$ the
graviscalar-pair invisible width can provide a 3\% to 20\% correction
to the direct-graviscalar-mixing invisible width. However, if $m_h$ is
substantially smaller than $M_D$, then the graviscalar pair width is
not important.  For example, for $\del=2$, $\mh=120\gev$ and
$M_D=500\gev$, $\Gamma(h'\to graviscalar~pairs)/\Gamma(h\to
graviscalar)<0.0015$ for $\xi<2$.  Therefore, in the following
analysis, where we will assume a light Higgs, we can safely neglect
the contribution to the invisible width from the decay into two
graviscalars and use the expression given by eq.~(\ref{invwidth}).

\section{MEASUREMENTS AT LHC}

For a Higgs boson with $m_h$ below the $WW$ threshold, the invisible
width causes a significant suppression of the LHC Higgs rate in the
standard visible channels.  For example, for $M_D=500\gev$ and
$\mh=120\gev$, $\Gamma(h\to graviscalar)$ is of order $25\gev$ already
by $\xi\sim 1$, {\it i.e.} far larger than the SM prediction of
$3.6\mev$.  Even when $m_h$ is greater than the $WW$ threshold,
Fig.~\ref{paircomp} shows that the partial width into invisible states
can be substantial even for $M_D$ values of several TeV; therefore,
for any given value of the Higgs boson mass, there is a considerable
parameter space where the invisible decay width of the Higgs boson
could be the first measured phenomenological effect from extra
dimensions.

Detailed studies of the Higgs boson signal significance, with
inclusive production, have been carried out by the {\sc
  Atlas}~\cite{:1999fq,unknown:1999fr} and {\sc Cms} ~\cite{cmsnote}
experiments. If 115~GeV $< \mh <$ 130~GeV, the $h \to \gamma \gamma$
channel appears to be instrumental for obtaining a $\ge 5 \sigma$
signal at low luminosity.  The $t \bar{t} h$, $h \to b \bar{b}$ and $h
\to ZZ^{*} \to 4~\ell$ channels also contribute, with lower statistics
but a more favorable signal-to-background ratio. Preliminary results
indicate that Higgs boson production in association with forward jets
may also be considered as a discovery mode. However, here the
background reduction strongly relies on the detailed detector
response.

In the ADD model, these results are 
modified by the appearance of an invisible decay 
width suppressing the Higgs signal in the standard
visible channels. 
Here, we fix $\mh=120\gev$ and perform a full scan of the ADD parameter 
space by varying $M_D$ and $\xi$ for different values of the number of extra dimensions $\delta$ and 
demonstrate that there are regions at high $\xi$ 
where the significance of the 
Higgs boson signal in the canonical channels drops below the 5~$\sigma$ threshold. 
However, the LHC experiments will also be sensitive to an invisibly decaying Higgs 
boson through $WW$-fusion production, 
with tagged forward jets. A detailed CMS 
study has shown that, with only 10~fb$^{-1}$, an invisible 
channel rate of 
$\Gamma_{inv}/\Gamma$=0.12-0.20 times the SM
$WW\to$Higgs production rate 
gives a signal exceeding the 5~$\sigma$ significance for 
120~GeV $< \mh  <$ 400~GeV~\cite{DiGirolamo:2001yv,cmsnote}. 
Given that the effective (from the sum
over the $h$ state and nearby degenerate states)
$WWh$ coupling is of SM strength,
this defines the region in the ADD parameter space
where the Higgs boson signal can be recovered 
through its invisible decay .

\begin{figure}[h]
\begin{center}
\begin{tabular}{c c}
\includegraphics[width=6.0cm,height=6.0cm]{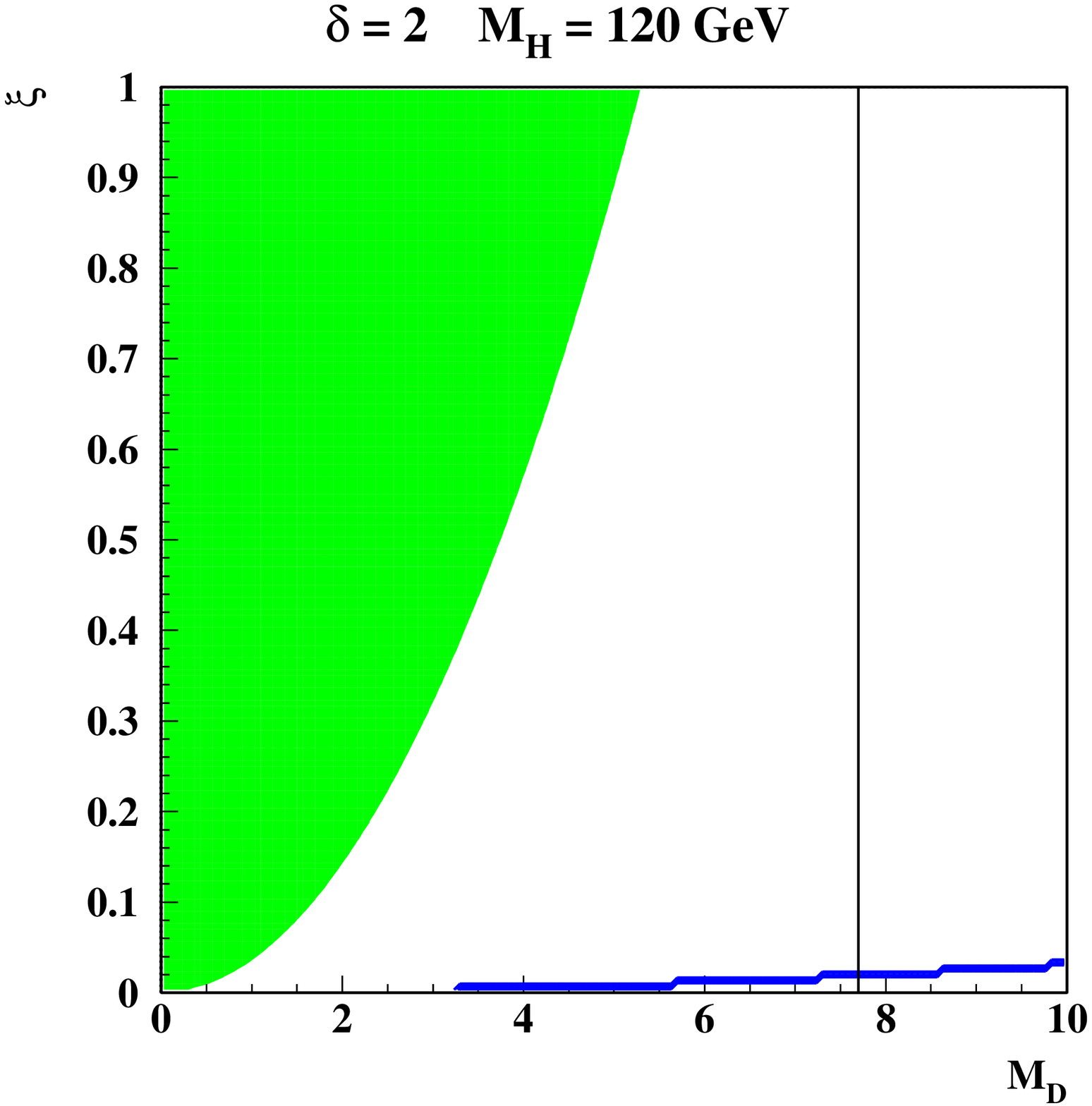} &
\includegraphics[width=6.0cm,height=6.0cm]{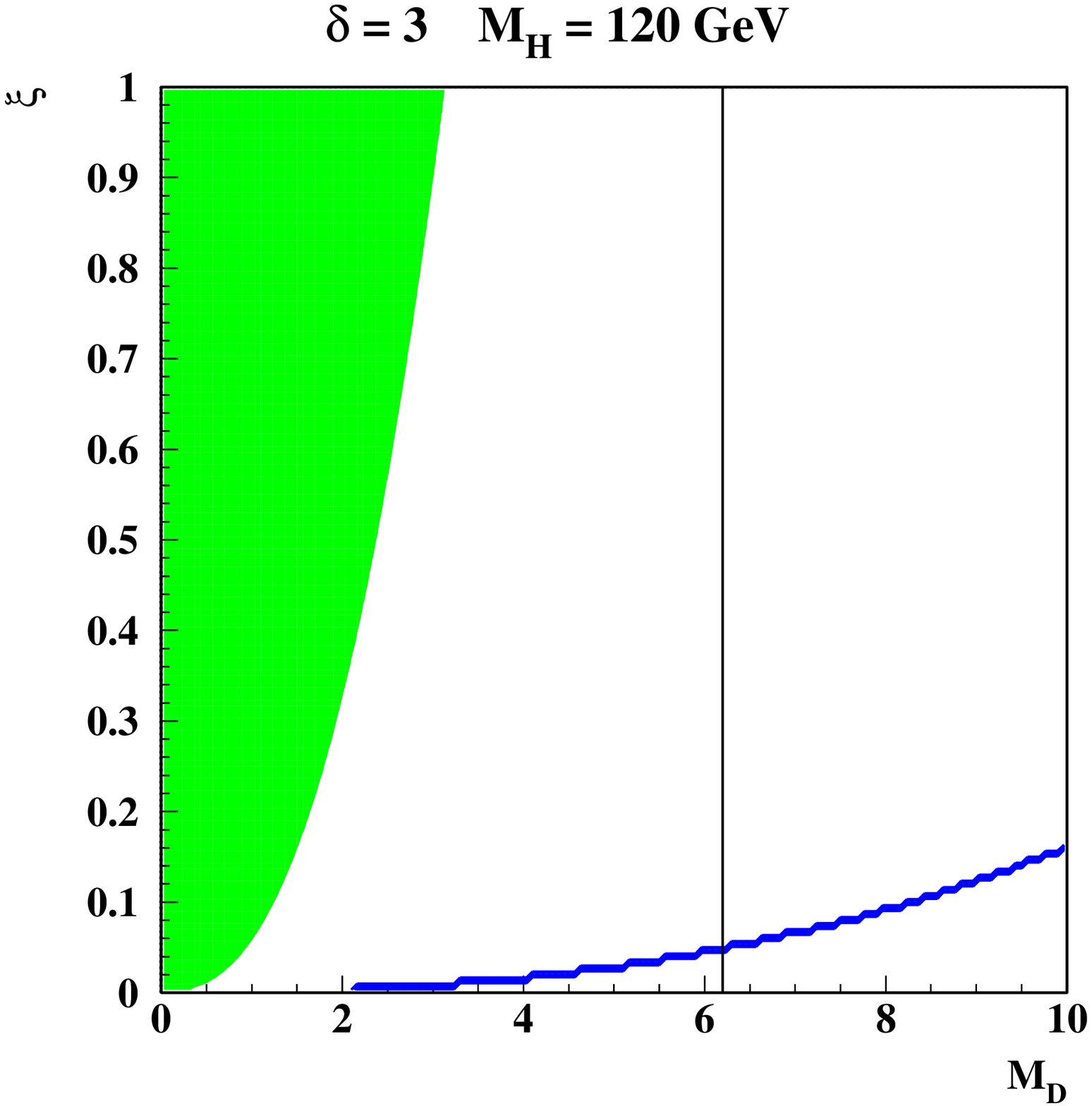}\\
\includegraphics[width=6.0cm,height=6.0cm]{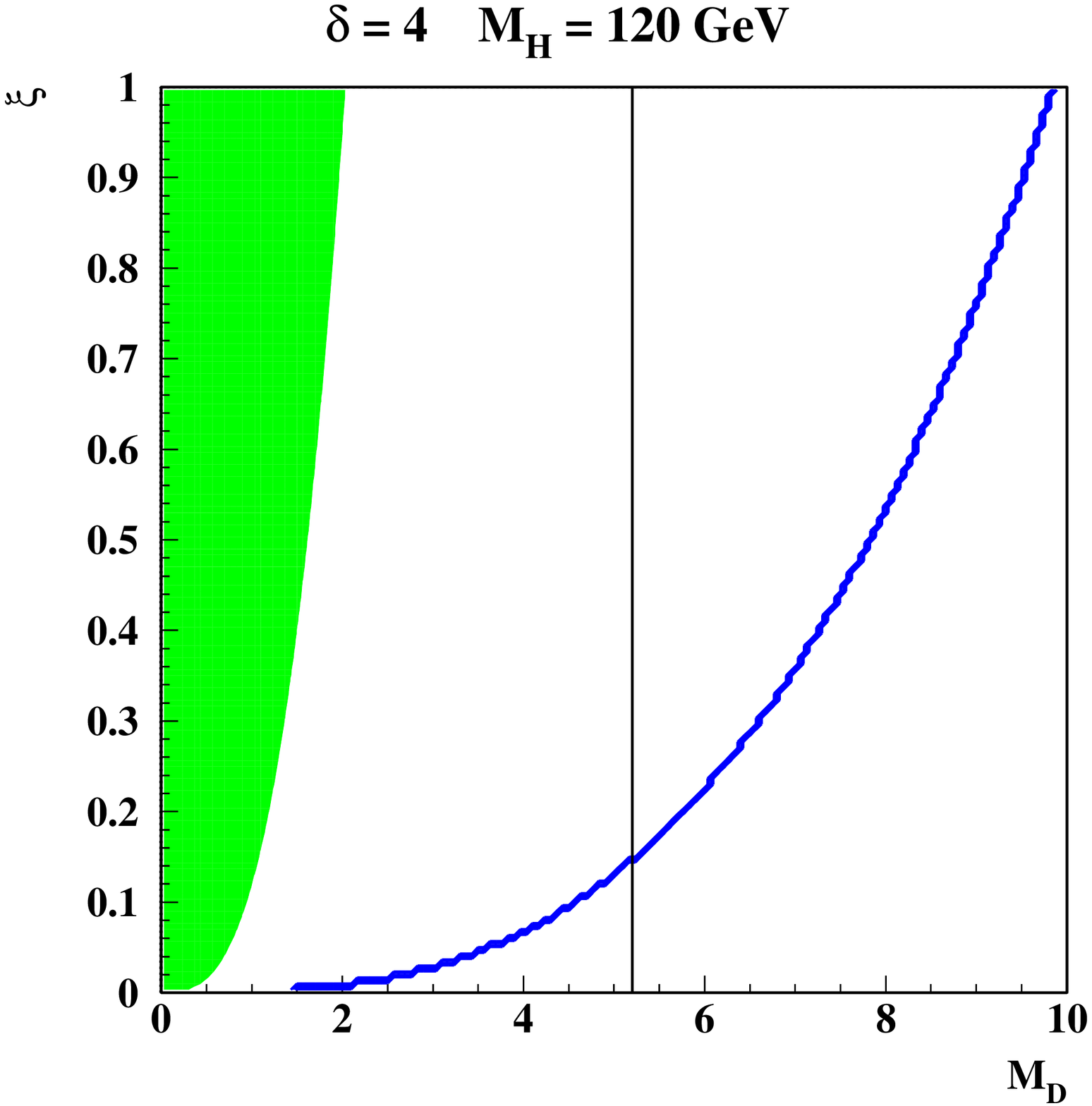}&
\includegraphics[width=6.0cm,height=6.0cm]{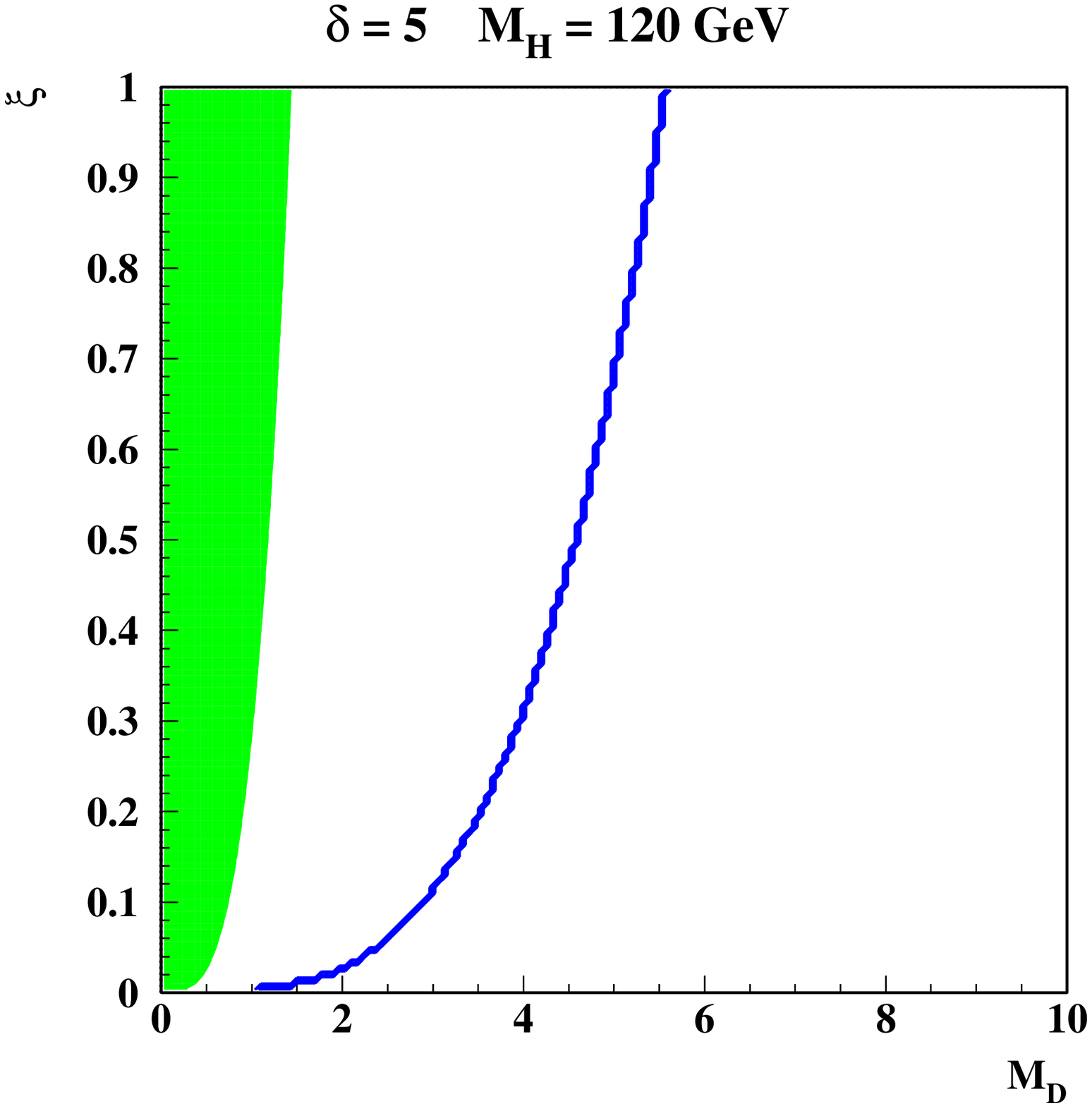}
\end{tabular}
\caption{Invisible decay width effects in the $\xi$ - $M_D$ plane for 
$M_H$ = 120~GeV. The green (grey) regions indicate where the Higgs signal 
at the LHC drops below the 5~$\sigma$ threshold for 30 $fb^{-1}$ of data. 
The regions above the blue (bold) line are the parts of the parameter space 
where the  invisible Higgs signal in the $WW$-fusion channel exceeds 5~$\sigma$
significance. The vertical lines show the upper limit on $M_D$  which can be 
probed by the analysis of jets/$\gamma$ with missing energy at the LHC. 
The plots are for different values of $\delta$: 2 (upper left), 3 (upper right)
4 (lower left), 5 (lower right).}
\label{figure1}
\end{center}
\end{figure}

Figure~ \ref{figure1} summarizes the results for specific choices of 
parameters. In the green (light grey) region, 
the Higgs signal in standard channels drops 
below the 5 $\sigma$ threshold  
with 30 $fb^{-1}$ of LHC data. But in the area above the
bold blue line the LHC search for invisible decays 
in the fusion channel yields a 
signal with an estimated significance exceeding 5 $\sigma$.
It is important to observe that, whenever the Higgs boson sensitivity is 
lost due to the suppression of the canonical decay modes, the invisible 
rate is large enough to still ensure detection through a dedicated analysis.

The analysis of Jet/$\gamma +$  missing energy is also sensitive to the ADD model 
over a range of the $M_D$ and $\delta$ parameters~\cite{Vacavant:2001sd}.
The invisible Higgs decay width appears to probe a parameter space up to, and beyond, 
that accessible to these signatures (see Figure~\ref{figure1}). Further, the sensitivity 
of these channels decreases significantly faster with $\delta$ compared to that of the 
invisible Higgs width if $\xi\sim 1$. 
Finally, it is interesting that, in the region where both 
signatures can be probed at the LHC, a combined analysis will provide a constraint 
on the fundamental theory parameters.

A TeV-class $e^+e^-$ linear collider will be able to further improve the determination 
of the Higgs invisible width. Extracting the branching fraction into invisible final 
states from the Higgsstrahlung cross section and the sum of visible decay modes affords 
an accuracy of order 0.2-0.03\% for values of the invisible branching fraction in the 
range 0.1-0.5. But the ultimate accuracy can be obtained with a dedicated analysis looking 
for an invisible system recoiling against a $Z$ boson in the $e^+e^- \to h Z$ process. 
A dedicated analysis has shown that an accuracy 
$0.04 < \delta {\mathrm{BR}}/{\mathrm{BR}} < 0.025$ can be obtained 
for $0.1 < {\mathrm{BR}} < 0.5$~\cite{schumacher}.
This accuracy would establish an independent constraint on the $M_D$, $\xi$ and $\delta$ 
parameters.

\vskip1cm
\noindent

\section*{ACKNOWLEDGEMENTS}
JFG and JDW are supported by the U.S. Department of Energy.

\bibliography{inv_proc_hepph}

\end{document}